\documentclass[aps,twocolumn,prd,showpacs,nofootinbib]{revtex4}
\usepackage{amsmath}
\usepackage{graphicx}
\usepackage{bm}
\usepackage{amssymb}
\usepackage{latexsym}
\usepackage{hyperref}
\usepackage{color}
\usepackage{appendix}
\hypersetup{ bookmarks=true,         
    unicode=false,          
    pdftoolbar=true,        
    pdfmenubar=true,        
    pdffitwindow=false,     
    pdfstartview={FitH},    
    pdftitle={My title},    
    pdfauthor={Author},     
    pdfsubject={Subject},   
    pdfcreator={Creator},   
    pdfproducer={Producer}, 
    pdfkeywords={keyword1} {key2} {key3}, 
    pdfnewwindow=true,      
    colorlinks=true,       
    linkcolor=red,          
    citecolor=cyan,        
    filecolor=magenta,      
    urlcolor=green,           
    linktocpage=true }

\newcommand{\ba}{\begin{eqnarray}}
\newcommand{\ea}{\end{eqnarray}}
\newcommand{\be}{\begin{equation}}
\newcommand{\ee}{\end{equation}}
\newcommand{\bea}{\begin{eqnarray}}
\newcommand{\eea}{\end{eqnarray}}
\newcommand{\beq}{\begin{equation}}
\newcommand{\eeq}{\end{equation}}
\newcommand{\beqar}{\begin{eqnarray}}
\newcommand{\eeqar}{\end{eqnarray}}
\newcommand{\beqars}{\begin{eqnarray*}}
\newcommand{\eeqars}{\end{eqnarray*}}
\newcommand{\bc}{\begin{center}}
\newcommand{\ec}{\end{center}}
\newcommand{\ben}{\begin{enumerate}}
\newcommand{\een}{\end{enumerate}}
\newcommand{\bit}{\begin{itemize}}
\newcommand{\eit}{\end{itemize}}
\newcommand{\bw}{\begin{widetext}}
\newcommand{\ew}{\end{widetext}}
\newcommand{\bcl}{\begin{columns}}
\newcommand{\ecl}{\end{columns}}
\newcommand{\ex}{\mbox{e}}
\newcommand{\dd}{\mbox{d}}

\newcommand{\ie}{\emph{i.e.~}}

\newcommand{\eg}{\emph{e.g.~}}

\newcommand{\Hu}{{\cal H}}

\newcommand{\Ka}{{\cal K}}

\newcommand{\GN}{G_{_\mathrm{N}}}

\newcommand{\mrm}[1]{\mathrm{#1}}
\newcommand{\mcl}[1]{\mathcal{#1}}

\newcommand{\ra}{\rangle}
\newcommand{\la}{\langle}
\newcommand{\lb}{\left(}
\newcommand{\rb}{\right)}
\newcommand{\lsb}{\left[}
\newcommand{\rsb}{\right]}
\newcommand{\lcb}{\left\{}
\newcommand{\rcb}{\right\}}
\newcommand{\nn}{\nonumber}

\def\setC{\mathbb{C}}

\def\setR{\mathbb{R}}

\newcommand{\GReCO}{${\cal G}\setR\varepsilon\setC{\cal O}$}

\begin{document}

\title{Non-Gaussianity excess problem in classical bouncing cosmologies}

\author{Xian Gao}
\email{gao@th.phys.titech.ac.jp}
\affiliation{Department of Physics, Tokyo Institute of Technology,
  2-12-1 O-Okayama, Meguro, Tokyo 152-8551, Japan}

\author{Marc Lilley}
\email{lilley@iap.fr}
\affiliation{Institut d'Astrophysique de Paris, UMR 7095-CNRS,
  Universit\'{e} Pierre et Marie Curie - Paris 6, 98bis Boulevard
  Arago, 75014 Paris, France,\\ \& \\ ACGC, Department of Mathematics and Applied
  Mathematics, University of Cape Town, Rondebosch 7701, South
  Africa}

\author{Patrick Peter}
\email{peter@iap.fr}
\affiliation{Institut d'Astrophysique de Paris, \GReCO, UMR 7095-CNRS,
  Universit\'{e} Pierre et Marie Curie - Paris 6, 98bis Boulevard
  Arago, 75014 Paris, France}

\date{\today}

\begin{abstract}
The simplest possible classical model leading to a cosmological bounce
is examined in the light of the non-Gaussianities it can generate.
Concentrating solely on the transition between contraction and expansion, 
and assuming initially purely Gaussian perturbations at the end
of the contracting phase, we find that the bounce acts as a source such
that the resulting value for the post-bounce $f_{_\mathrm{NL}}$ may
largely exceed all current limits, to the point of potentially casting
doubts on the validity of the perturbative expansion.  We conjecture
that if one can assume that the non-Gaussianity production depends
only on the bouncing behavior of the scale factor and not on the
specifics of the model examined, then many realistic models in which a
nonsingular classical bounce takes place could exhibit a generic
non-Gaussianity excess problem that would need to be addressed for
each case.
\end{abstract}


\maketitle

\section*{Introduction}

The recently released P{\footnotesize LANCK} data
\cite{Ade:2013uln,Ade:2013zuv} have set new standards as far as
cosmological modeling is concerned, imposing very tight constraints on
early universe physics \cite{Mukhanov:2005sc,PeterUzan2009} and
discriminating \cite{Martin:2013nzq,Linde:2014nna} among numerous
inflationary theories \cite{Martin:2014vha}. Bouncing cosmologies are
among very few possibly viable alternatives to inflationary cosmology
(see \cite{Battefeld:2014uga} for a review). This being said, the only
relevant bouncing models worth investigating
\cite{Peter:2008qz,Brandenberger:2012zb,Battefeld:2014uga}, are those
that are able to reproduce the observed power spectra, both scalar and
tensorial. In turn, these
models have to face the most serious cosmological constraint to date,
namely that imposed by the smallness of non-Gaussianities
\cite{Ade:2013ydc}. Whether or not generic bouncing models can
successfully pass this test will decide on their viability.  To a
large extent, the non-Gaussianity parameter $f_{_\mathrm{NL}}$ does
not depend on the actual spectrum of first order perturbations, and is
thus also independent of their initial conditions. This makes it an
invaluable tool to assess the viability of any cosmological model.

The purpose of the present paper is to demonstrate, by means of an
explicit calculation, itself drawing heavily on the ones detailed in
Ref.~\cite{Gao:2014hea}, that the non-Gaussianity produced 
during the transition from contraction to expansion, and thus by the bounce
itself, may far exceed existing
contraints on $f_{_\mathrm{NL}}$. Recalling that canonical single field
slow-roll inflation naturally predicts small $f_{_\mathrm{NL}}$, our findings would tend to favor
the inflationary paradigm by disqualifying one of its few alternatives.

The particular category of model studied in this paper is that for which the matter 
content is in the form of a strictly positive energy scalar field.  
The presence of a negative energy component being crucial for the obtention of a 
bounce, we take the spatial curvature to be positive, so that it acts as an {\it effective} 
negative energy component. While it is true that many bouncing models are constructed with a vanishing or negligible
spatial curvature contribution, they necessarily involve other types
of negative energy fields, which may cause serious instabilities, and
hence also potentially produce large amounts of non-Gaussianities.
Therefore, although the results which we 
present below apply, strictly speaking,
to nonsingular bouncing models dominated at the bounce by the
positive spatial curvature term in the Friedmann equation, and for
which General Relativity (GR) is valid all along, we conjecture that
it could apply to a much wider set of similarly nonsingular models,
hence raising a possibly generic problem with bouncing cosmologies.  
Note that we do not consider singular bounces for which GR does not
apply throughout as no reasonable prediction can be made in such contexts 
without an explicit calculation within the framework of an (as of yet still unknown) theory of quantum gravity.

\section{Theoretical framework}

We start from the GR action (we work in natural units in which
$8\pi\GN=c=\hbar=1$),
\begin{equation}
S=\int \dd^4x\sqrt{-g} \left( -R+\mathcal{L}_\mathrm{mat}\right)
\end{equation}
where $\mathcal{L}_\mathrm{mat}$ describes the matter content
and $R$ is the Ricci scalar derived from the metric tensor
$g_{\mu\nu}$. The metric itself is chosen to be that of a
perturbed Friedmann-Lema\^itre 
line element, given in Poisson gauge\footnote{This gauge is known to
introduce potentially large and unphysical effects. The quantities
calculated below however, being the ratios of spectra of first order perturbations should not be plagued by this problem.} by
\begin{equation}
\dd s^{2}=a^{2}\left(-\ex^{2\Phi}\dd\eta^{2}
+\ex^{-2\Psi}\gamma_{ij} \dd x^{i}\dd x^{j}\right),
\label{metric_pert_poisson}
\end{equation}
where
\[
\gamma_{ij}=\lb 1+\frac14\Ka \delta_{mn}x^m x^n\right)^{-2}\delta_{ij}
\]
is the background spatial metric which we
take to be of constant positive curvature ($\Ka=1$).  The fields
\[
\Psi=\sum_i\frac{\Psi_{(i)}}{i!} \ \ \ \hbox{and} \ \ \ 
\Phi=\sum_i\frac{\Phi_{(i)}}{i!}
\]
are the Bardeen potentials up to
arbitrary order in perturbations and encode the scalar cosmological
fluctuations in the metric.  Note that here, one has, at first order,
$\Psi_{(1)}=\Phi_{(1)}$.

The background metric, \ie that obtained in the limit $\Psi,\Phi\to 0$,
satisfies the Friedmann equations
\begin{equation}
\Hu^2 + \Ka = \frac{1}{3} a^2 \rho,
\label{FriedBack}
\end{equation}
where $\rho$ is the fluid energy density and the conformal Hubble
rate is $\Hu \equiv a'/a$, a prime meaning a derivative with respect to.~the conformal
time $\eta$. The normalized energy density is defined through
$\Omega\equiv \rho a^2/(3\Hu^2)$, and one may associate to the spatial curvature term
$\Ka$ a normalized energy density in a similar way through $\Omega_\Ka\equiv\Ka/\Hu^2$.

We set $\Ka=1$ for two reasons.  

First, as stressed in the
Introduction, the obtention of a bounce requires the presence of an
effectively negative energy component.  Positive spatial curvature is
its simplest incarnation.  It is free of the instabilities that may for instance
result from the introduction of ghost fields, and less speculative than
for example the Galileon/ghost condensate implementation
(see, \eg Ref.~\cite{Cai:2013vm} and references therein). Whether
or not this latter implementation exhibits the large non-Gaussianity
problem discussed in the present paper is still a matter of debate. 

Second, spatial
curvature is identically zero only in the special and entirely
implausible situation where $\Omega_\Ka=0$ strictly. We would argue that $\Omega_\Ka=0$ can only be
 the result of extreme fine-tuning or occurs in specific
theoretical frameworks (\eg brane inflation in superstring theory where spatial
flatness and isotropy are protected by symmetry).  In general, in any realistic
cosmology, $\Omega_\Ka\ne 0$, with current observational constraints
to some extent favoring a slightly closed universe with $\Ka=1$
\cite{Ade:2013zuv}.  Furthermore, at the bounce, the Hubble parameter
$\Hu$ being equal to zero, it is the balance between the spatial
curvature term and the energy contents of the cosmology which
determines the dynamics.  Under general conditions, spatial curvature
can thus by no means be assumed negligible at the bounce point
when otherwise only positive energy density components are present. In the
case of a model that relies on a ghost condensate or some other effectively negative energy
density component, the negligibility of the spatial curvature term can only be invoked {\it a posteriori}, 
\ie if an explicit calculation of $a_{_\mathrm{B}}$, the scale factor at the bounce, demonstrates that it is indeed negligible.\footnote{Here and in what follows, the 
subscript {\footnotesize  ``B''} denotes a quantity evaluated at
the time of the bounce.}

Although non-negligible at the bounce, the spatial curvature at 
late times can easily be made to agree with current limits on $\Omega_\Ka$.  
This can be achieved in two different ways. The first is
the existence of a phase of inflation following the bounce
\cite{Falciano:2008gt,Lilley:2011ag}. The second is the existence of a
phase of deflation prior to the bounce \cite{Gasperini:1993hu} with
the added requirement that the bounce be close to symmetric (see
\cite{Battefeld:2014uga}).

\begin{figure}[h!]
\begin{center}
\includegraphics[width=8.7cm]{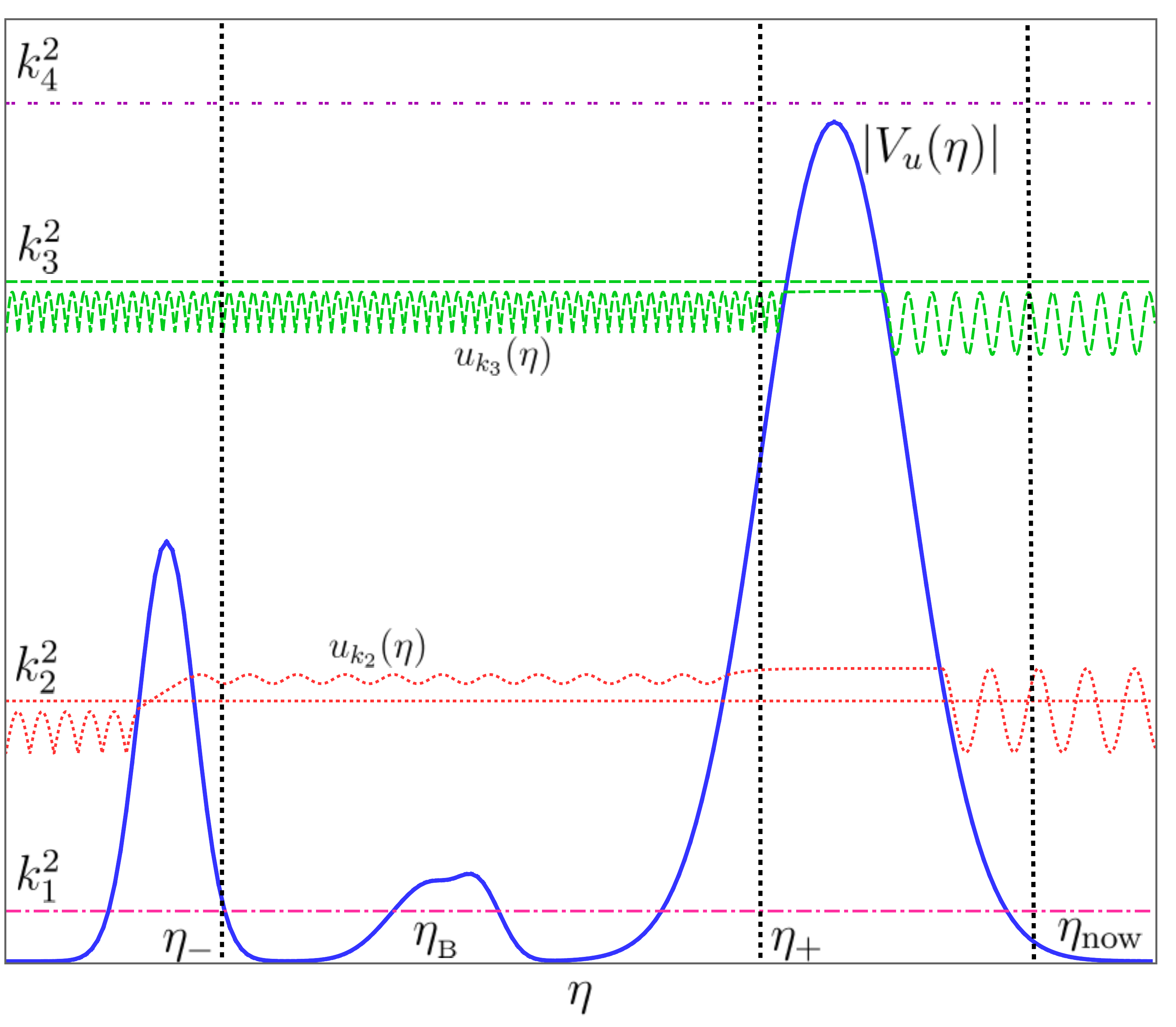}
\caption{Prototypical potential $V_u(\eta)$, and wave number squared
  (see \cite{Martin:2003sf,Falciano:2008gt,Lilley:2011ag} for explicit
  examples). The bounce itself occurs between $\eta_-$ and
  $\eta_+$. At the level of the two-point statistics, small scale
  perturbations (\eg those of wave number $k_4$) remain unaffected,
  while long wavelength perturbations ($k_1$, $k_2$ or $k_3$) can be
  spectrally modified in different ways.  For illustrative purposes,
  the time evolution of two modes, $u_{k_2}$ and $u_{k_3}$ is also
  shown. As shown in this paper, the bounce produces large
  non-Gaussianities for any $\lcb k_1,k_2,k_3\rcb$ configuration. The
  first peak before $\eta_-$ might represent an initial source for
  primordial perturbation enhancement, as \eg a matter or ekpyrotic contraction,
  while the second peak, after $\eta_+$, could be understood as an
  inflationary stage subsequent to the bounce. Although during both
  these phases, further non-Gaussianity could be produced, we
  restrict attention here to the seemingly more harmless period
  between $\eta_-$ and $\eta_+$, \ie the bounce itself.}
\label{potns}
\end{center}
\end{figure}

We now restrict attention to the specific case for which the matter consists
in a single scalar field $\phi$ with
a canonical kinetic term and evolving in a potential $V(\phi)$. We therefore have
\begin{equation} S=-\int \dd^4x\sqrt{-g}\lsb
R+\lb\partial\phi\rb^2+V(\phi)\rsb.
\label{eq:action}
\end{equation}
At the level of first order perturbations, introducing the variable
$u\propto a\Psi_{(1)}/\phi'$ and its Fourier modes, defined by $\Delta
u_{\bm k} = -k^2 u_{\bm k}$, one finds \cite{Martin:2003bp}
\begin{equation}
u''_{\bm k}+\lsb k^2-V_u(\eta)\rsb u_{\bm k}=0,
\end{equation}
where the potential $V_u(\eta)$ is sketched in Fig.~\ref{potns},
drawing on the specific functional shapes of $V_u(\eta)$ obtained in
previous works on the same model
\cite{Martin:2003sf,Falciano:2008gt,Lilley:2011ag}.  As
shown in the figure, a typically asymmetric bouncing phase occurs at
$\eta_{_\mrm B}$
  and is generically preceded and followed by peaks in
the potential with model-dependent amplitudes and widths.  The peak
that occurs prior to the bounce follows a regime in which $V_u$
vanishes, in such a way that unambiguous vacuum initial conditions can
be set.  In contrast with what happens in inflation, for which modes
cross the potential only once (\eg the mode with wave number labeled
$k_3$ in Fig.~\ref{potns}), in a
bouncing cosmology, modes may cross the potential three or more times
(\eg modes with wave numbers $k_1$ or $k_2$ in Fig.~\ref{potns}). 
The primordial spectrum is
therefore modified for wave numbers $k_1$, $k_2$, with possibly
superimposed oscillations \cite{Falciano:2008gt,Lilley:2011ag} and, as
will be shown below, the amplitude of the three-point function of
cosmological perturbations generated by the bounce for such
scales can consequently be very large \cite{Gao:2014hea}.

At this stage in the discussion, it is possible to make one more
argument, at the level of first order perturbations, towards
the genericity of the analysis presented here, and its nonspecificity
to spatial curvature dominated bounces. The shape of the
potential $V_u(\eta)$ was discussed in detail in
Ref.~\cite{Martin:2003sf}.  In a Taylor expansion in the vicinity of
the bounce, the potential for the rescaled Bardeen variable $u$ at the
bounce is characterized by its width and height, each given by Eqs
(52) and (53) of that paper.  From these equations, it is easily seen
that the potential depends mainly on the kinetic term $(1/2)(\phi')^2$
and on the logarithmic derivatives of $V(\phi)$.  It does not depend
crucially on spatial curvature.  In fact as shown in
Refs~\cite{Falciano:2008gt,Lilley:2011ag}, spatial curvature enters in the
potential of first order perturbations through a constant term equal to $4$.
It can also be noted that taking the limit $\Ka\to 0$ in the final
results obtained below yields exactly the same conclusions.

\section{Modeling the bounce}

In this paper, we focus on the calculation of the amount of
non-Gaussianity produced by the bouncing phase only. It is thus
sufficient for our purpose to expand the scale factor around the
bounce in powers of conformal time $\eta$,
\begin{equation}
\frac{a}{a_0}=1+\frac{1}{2}\left(\frac{\eta}{\eta_\mathrm{c}}\right)^2+
\lambda_3\left(\frac{\eta}{\eta_\mathrm{c}}\right)^3+
\frac{5(1+\lambda_4)}{24}\left(\frac{\eta}{\eta_\mathrm{c}}\right)^4 +\cdots,
\label{eq:scalefactor}
\end{equation}
where $\eta_{\mrm{c}}$ is the characteristic time scale of the bounce,
and to compute the production of non-Gaussianity between an initial
spatial hypersurface at time $\eta_-$ satisfying $-\eta_{\mrm
  c}\lesssim \eta_-<0$ and a final spatial hypersurface at time
$\eta_+$ satisfying $0\lesssim \eta_+<\eta_{\mrm c}$. In
Eq.~(\ref{eq:scalefactor}), we have set the bounce conformal time
$\eta_{_\mrm B}=0$ for convenience. The two additional constants $\lambda_3$
and $\lambda_4$ parametrize deviations from a de Sitter bounce at
cubic and quartic order in $\eta$ respectively while $\eta_{\mrm c}$
is an overall deviation in the bouncing time scale from the de Sitter
bouncing time scale.

At the level of the background cosmology, introducing the parameter
$\Upsilon=\phi_{_\mathrm{B}}'^2/2$, one
may use the Einstein equations to express the bouncing time scale as
$\eta_\mathrm{c}=(1-\Upsilon)^{-1/2}\ge 1$. Two additional parameters
$\varepsilon_V=(V_{,\phi}/V)|_{_\mathrm{B}}$ and
$\eta_V=(V_{,\phi\phi}/V)|_{_\mathrm{B}}$ can be related to $\Upsilon$,
$\lambda_3$ and $\lambda_4$ in Eq.~(\ref{eq:scalefactor}) through the
Einstein equations, with the de Sitter bounce being recovered in the
limit $\Upsilon\to0$
\cite{Martin:2003sf,Falciano:2008gt,Gao:2014hea} (recall that 
one expects the de Sitter solution to be an attractor for this dynamical
system). In terms of $\Upsilon$, $\varepsilon_V$ and $\eta_V$, the bounce is seen
to be controlled by the kinetic energy of $\phi$ and the flatness of
the potential $V(\phi)$.

The equation of motion for the Fourier modes of perturbation
at the $i^\mathrm{th}$ order reads
\begin{equation}
\mcl D\,\Psi_{(i)}=\mcl S \lsb \Psi_{(i-1)}\rsb,
\label{eq:eom}
\end{equation}
where
\[
\mcl D=\partial_{\eta}^2+F\lb
\eta\rb\partial_{\eta}+k^2+W(\eta)
\]
(the subscript ``$\bm k$'' on the
modes is not written explicitly but is instead implicitly assumed for notational
simplicity), with 
\[
F(\eta)=2\lb \mcl H-\frac{\phi''}{\phi'}\rb
\]
and
\[
W(\eta)=2\lb \mcl H'-\mcl H\frac{\phi''}{\phi'}-2\mcl K\rb.
\]
The source term
$\mcl S\lsb \Psi_{(i-1)}\rsb$ is vanishing for $i=1$ and its explicit
form for $i=2$, not essential for the present discussion, was computed
in \cite{Gao:2014hea} and depends on quantities computed
at all previous orders.

\section{Non-Gaussianities}

The series solution of Eq.~(\ref{eq:eom}) for $\Psi_{(1)}$ up to order
$\eta^2$ can be written in terms of two mode functions $v_1(k,\eta)$
and $v_2(k,\eta)$ normalized at the prebounce time $\eta_-$ (see Fig.~\ref{potns})
in such
a way that $v_1(k,\eta_-)=1$,
$v_1'(k,\eta_-)=0$, $v_2(k,\eta_-)=0$ and $v_2'(k,\eta_-)=1$
\cite{Gao:2014hea}.  In this basis, the initial conditions
are given in terms of a set of random variables $\hat x_a\equiv
\left\{ \Psi_{(1)}(\eta_-),\Psi_{(1)}'(\eta_-)\right\}$ providing the
initial conditions of the first order perturbation and its time
derivative on the initial spatial hypersurface. As we are interested
in the amount of non-Gaussianity produced during the bouncing phase, we
shall assume that the variables $\hat x_a$ follow Gaussian
statistics. The $\hat x_a$ in turn define a spectral matrix $\bm P$ at
$\eta_-$ by $\la\hat{x}_{a}\lb\bm{k}_1\rb\hat{x}_{b}\lb\bm{k}_2\rb\ra
=\delta_{\bm{k}_1\bm{k}_2} P_{ab}\lb k\rb$, where the indices
$a,b$ represent either $\Psi$ or $\Psi'$. It is important to note that, in general,
and in contrast to the more usual inflationary case, all four entries
in $\bm P$ are necessary to calculate the amount of non-Gaussianity
produced by the bouncing phase since we cannot assume the mode
to have reached the constant super-Hubble value which is characteristic
of the more usual inflationary evolution.  Note also that the background
spacetime being of constant positive curvature, all calculations are
performed on the three-sphere $\mathbb{S}^3$ and the wave vectors
consist in three integer numbers: $n>1$, giving the amplitude
$k^2=n(n+2)$; $\ell>0$; and $m\in [-\ell,\ell]$, while
$\delta_{\bm{k}_1\bm{k}_2}$ is the product of three Kronecker delta
functions $\delta_{n_1n_2}$, $\delta_{\ell_1\ell_2}$, and
$\delta_{m_1m_2}$.

\begin{figure*}[t]
\begin{center}
\includegraphics[width=8.9cm]{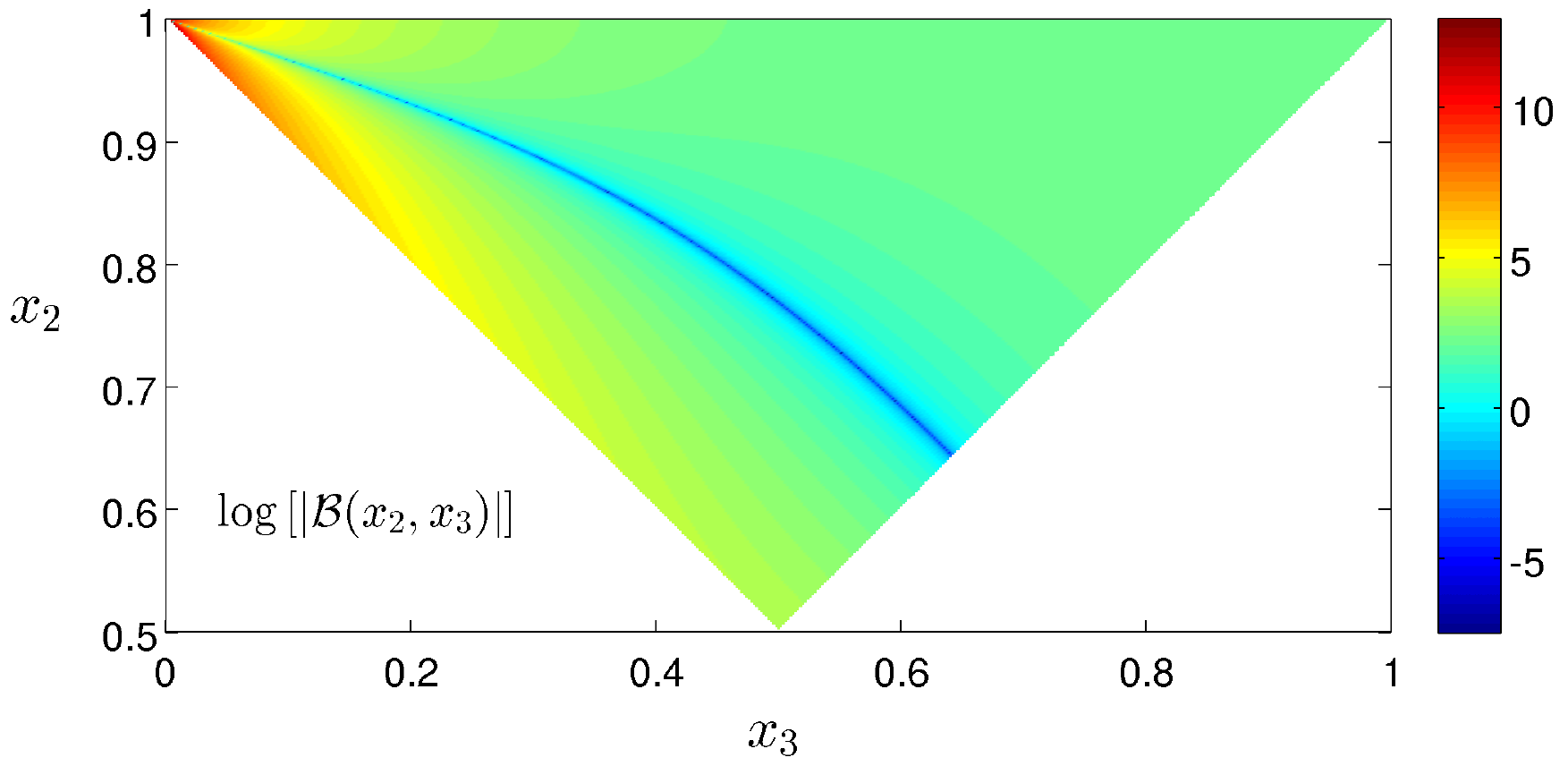}
\includegraphics[width=8.9cm]{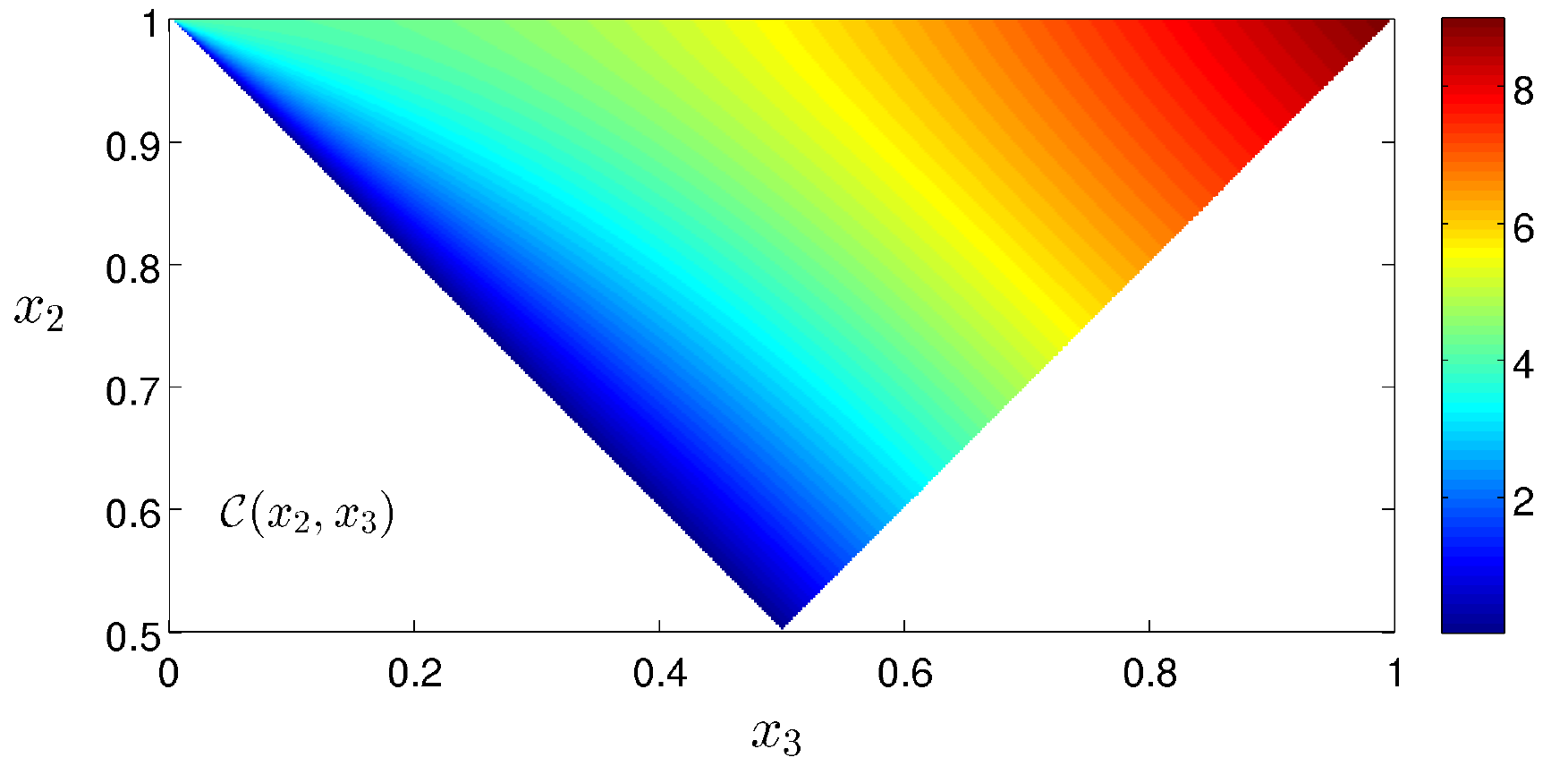}
\includegraphics[width=8.9cm]{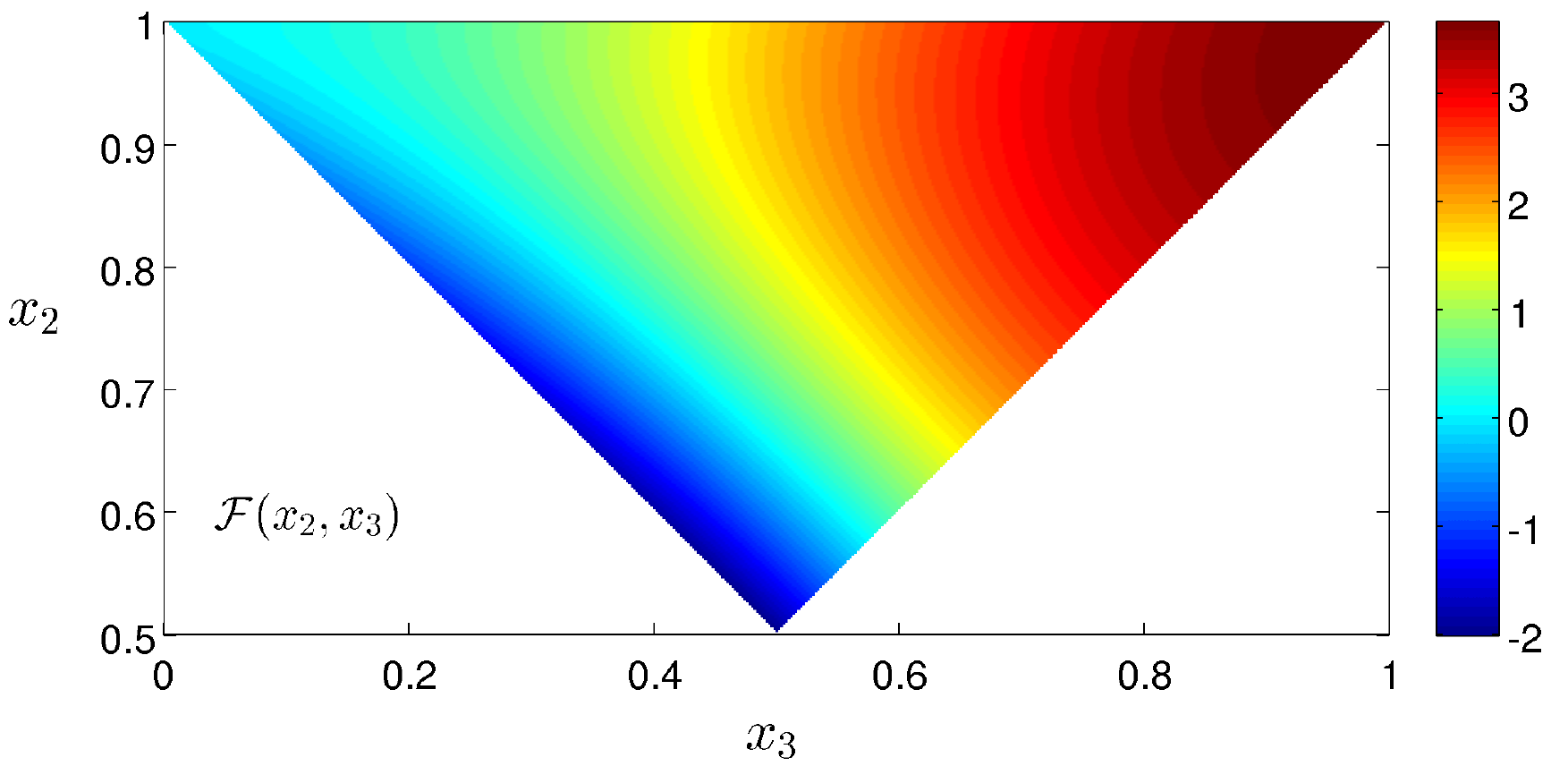}
\includegraphics[width=8.9cm]{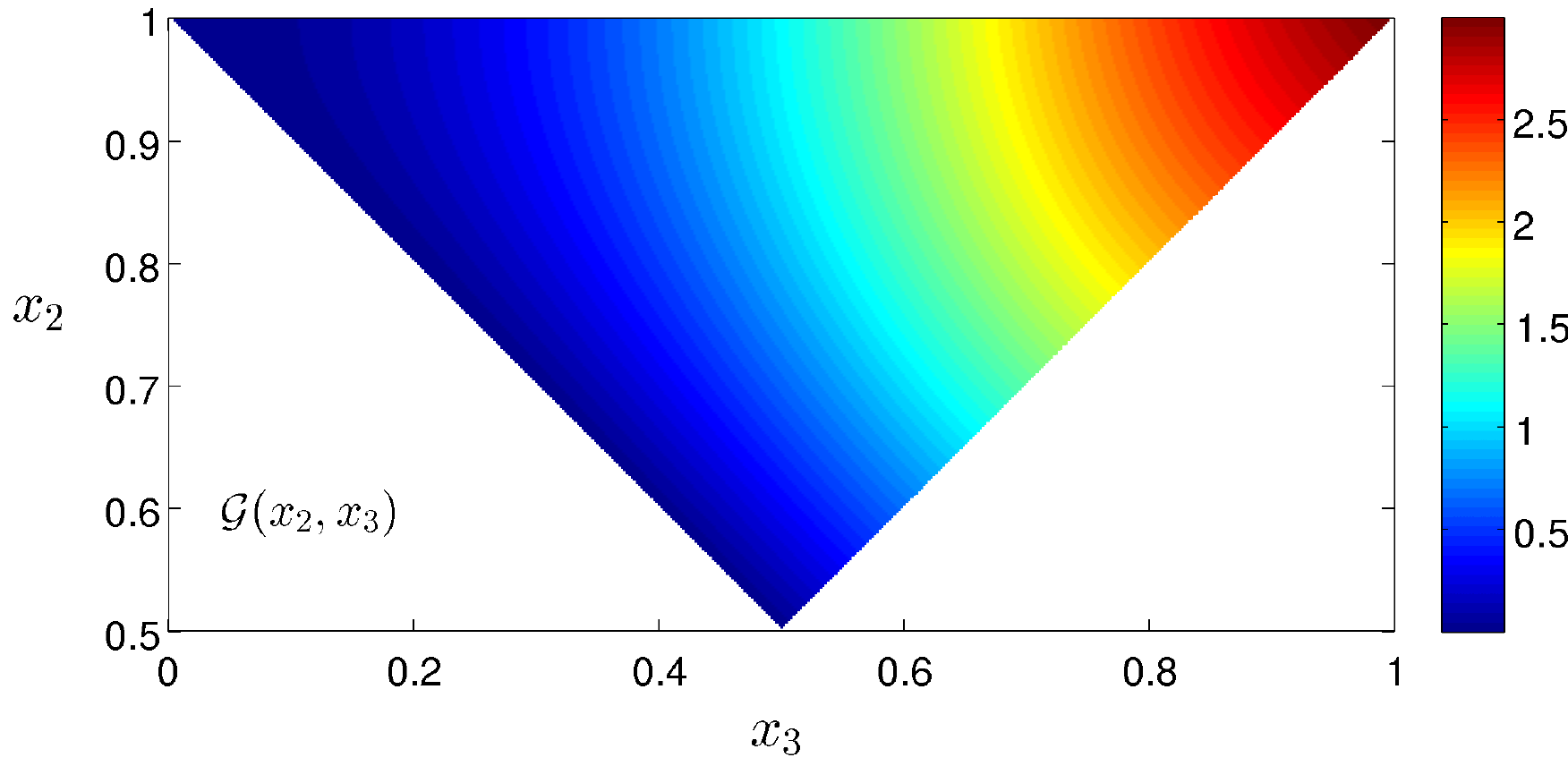}
\caption{Shape functions derived from Eq.~(\ref{eq:fnl}) showing the
relative contributions of the various possible non-Gaussian
configurations. Top figures: $\log(|\mcl B|)$ (left) and $\mcl C$ (right) 
obtained assuming $K_1 (k_i) K_{1,2}(k_j)\simeq K_3(k_1,k_2,k_3)$. Lower
figures: almost scale-invariant
``frozen'' state approximation ($P_{\Psi'\Psi'}\ll P_{\Psi\Psi'}\ll P_{\Psi\Psi}
\propto k^{n_\mathrm{s}-4}$ with $n_\mathrm{s} = 0.9603$);
the figures  are the shape functions obtained by combining Eqs~(20) to (24).
In the  figures on the left, $f_{_\mrm{NL}}\propto \Upsilon^{-1}$. 
In the figures on the right,  $f_{_\mrm{NL}}\propto (k_1^2/\Upsilon)$.
In all four figures, $x_2=k_2/k_1$ and $x_3=k_3/k_1$. The differences
in the amplitude as a function of the configuration $\lcb k_1,
k_2,k_3\rcb$ highlight the dependence of the shape function on the
details of $\bm P$.}
\label{figure2}
\end{center}
\end{figure*}

The bispectrum $\mcl B_{\Psi}$ produced during the bouncing phase (\ie, in the
interval $\eta_-$ to $\eta_+$, as shown on Fig.~\ref{potns})
is defined through the three-point function of the perturbation $\Psi$, evaluated at $\eta_+$
\cite{Gao:2014hea},
\begin{equation} \la
\Psi_{\bm{k}_{1}}\Psi_{\bm{k}_{2}}\Psi_{\bm{k}_{3}}\ra =\frac{1}{2}\mcl
G_{\bm{k}_{1}\bm{k}_{2}\bm{k}_{3}}\mcl B_{\Psi}\lb k_{1},k_{2},k_{3}\rb,
\end{equation}
where $\mcl G_{\bm k_1\bm k_2\bm k_3}$ is a geometrical form factor
generalizing the flat case $\delta\lb\bm k_1+\bm k_2+\bm k_3\rb$ to
$\mathbb{S}^3$; it is given by an integral over the product of three
hyperspherical harmonics. The bispectrum is used to define the
nonlinearity parameter $f_{_\mrm{NL}}$, obtained by expressing the
non-Gaussian signal in terms of the sum of squares of the two-point
functions for wave numbers $k_1$, $k_2$ and $k_3$ through

\begin{widetext}
  \begin{equation}
    \mcl B_{\Psi}(k_{1},k_{2},k_{3})=\frac{6}{5}f_{_\mathrm{NL}} \lsb
P_{\Psi\Psi}(k_{1})P_{\Psi\Psi}(k_{2})+P_{\Psi\Psi}(k_{2})P_{\Psi\Psi}(k_{3})+
P_{\Psi\Psi}(k_{3})P_{\Psi\Psi}(k_{1})\rsb.
  \end{equation}

\noindent Using the results obtained in \cite{Gao:2014hea},
we now calculate $f_{_\mathrm{NL}}$ at leading order in
$\Upsilon$, $\varepsilon_V$ and $\eta_V$ and in the limit of large
wave numbers $k$.  This latter assumption is justified because the
range of observationally accessible physical wave numbers today is
$10^{-2}h\,\mrm{Mpc}^{-1}\lesssim k_{\mrm{phys}} \lesssim
10^3h\,\mrm{Mpc}^{-1}$ and corresponds to a range of comoving
wave numbers $10^2\lesssim k\lesssim 10^8$ for a conservative value
$\Omega_{\mcl K}\sim 10^{-2}$ \cite{Ade:2013zuv} (P{\footnotesize LANCK}
latest
results indicating $\Omega_{\mcl K} \lesssim 5\times 10^{-3}$). We find
\begin{align}
f_{_\mathrm{NL}} &= -\frac{5(k_1+k_2+k_3)}{3\Upsilon\displaystyle
  K_3\lb
  k_1,k_2,k_3\rb}\,\left[\displaystyle\prod_{\sigma(i,j,\ell)}\lb
  k_i+k_j-k_{\ell}\rb\right]\lcb\displaystyle\sum_{\sigma(i,j,\ell)}
\frac{K_1(k_i)K_1(k_j)}{k_{\ell}^2}-4\lsb
\frac{K_1(k_i)K_2(k_j)}{k_j^2k_{\ell}^2}+\frac{K_1(k_j)K_2(k_i)}{k_i^2
  k_{\ell}^2}\rsb\rcb\nn\\ & +\frac{5}{3\Upsilon\displaystyle K_3 \lb
  k_1,k_2,k_3\rb} \sum_{\sigma(i,j,\ell)}\lsb
\frac{7}{3}+\frac{2}{3}\lb
\frac{k_i^2+k_j^2}{k_{\ell}^2}\rb-3\lb\frac{k_i^2-k_j^2}
     {k_{\ell}^2}\rb^2\rsb K_1(k_i)K_1(k_j)+\cdots,
\label{eq:fnl}
\end{align}
where the dots denote subleading terms in inverse powers of $k$ and
higher order in $\Upsilon$, $\varepsilon_V$ and $\eta_V$. In Eq.~(\ref{eq:fnl}),
the relevant functions of the initial spectra are
\begin{equation}
K_1(k)=6P_{\Psi\Psi}(k)+7P_{\Psi\Psi'}(k)+2P_{\Psi'\Psi'}(k)\,, \qquad
K_2(k)=7P_{\Psi\Psi}(k)+11P_{\Psi\Psi'}(k)+4P_{\Psi'\Psi'}(k)\,,
\end{equation}
and
\begin{align} K_3(k_1,k_2,k_3)&=
81\sum_{\sigma(i,j)}P_{\Psi\Psi}(k_i)P_{\Psi\Psi}(k_j) +108
\sum_{\sigma(i,j)}P_{\Psi\Psi}(k_i)P_{\Psi\Psi'}(k_j)+36
\sum_{\sigma(i,j)}P_{\Psi\Psi}(k_i) P_{\Psi'\Psi'}(k_j)+\nn\\ &
144\sum_{\sigma(i,j)}P_{\Psi\Psi'}(k_i)P_{\Psi\Psi'}(k_j)+ 48
\sum_{\sigma(i,j)} P_{\Psi\Psi'}(k_i)P_{\Psi'\Psi'}(k_j)+16
\sum_{\sigma(i,j)}P_{\Psi'\Psi'}(k_i)P_{\Psi'\Psi'}(k_j),
\label{Kdefs}
\end{align}
so in the general case, the non-Gaussianity parameter
$f_{_\mathrm{NL}}$ depends not only on the spectrum of curvature
perturbations $P_{\Psi\Psi}$ but also on that of its time derivative
$P_{\Psi'\Psi'}$ as well as on the cross spectrum $P_{\Psi\Psi'}$, both
usually assumed irrelevant in the usual inflationary framework.
\end{widetext}

\noindent In Eqs.~(\ref{eq:fnl}) and (\ref{Kdefs}), the sums and
products are taken over all possible permutations of $i$, $j$ and
$\ell$ with $\sigma(i,j,\ell)$ denoting $(i,j,\ell) \in
\{(1,2,3),(1,3,2),(2,3,1)\}$, and $\sigma(i,j)$ denoting $(i,j) \in
\{(1,2),(1,3),(2,3)\}$. In the equilateral ($k_1=k_2=k_3=k$) and
squeezed ($k_i=k_j=k$ and $k_{\ell}=p\ll k$) configurations and at
leading order, Eq.~(\ref{eq:fnl}) simplifies to
\begin{eqnarray}
f_{_\mrm{NL}}^{\mrm{equi}}&=&-\frac{15k^2}{\Upsilon}
\frac{K_1^2(k)}{K_3(k,k,k)}, \label{fNLequiv}
\\ f_{_\mrm{NL}}^{\mrm{sq}}&=&-\frac{20k^2}{3\Upsilon}
\frac{K_1^2(k)+K_1(k)K_1(p)}{K_3(k,k,p)},
\label{fNLsq}
\end{eqnarray}
so that the non-Gaussianity parameter is of order $k^2/\Upsilon$. In
the folded configuration ($k_2 = k_3 = \frac12 k_1$), the first
nonvanishing term is given in the second line of Eq.~(\ref{eq:fnl})
and simplifies to
\begin{equation}
f_{_\mrm{NL}}^{\mrm{fold}}=\frac{40}{9\Upsilon}\frac{K_1(k)\lsb K_1(k)
-16K_1(2k)\rsb}{K_3(k,k,2k)}\,.
\label{fNLfold}
\end{equation}
The square of the wave number does not appear in the numerator of
Eq.~(\ref{fNLfold}) so that the folded configuration is in general subdominant relative to the equilateral and squeezed configurations.

\section{Discussion}

Given that the matrix $\bm P$ is unknown, the $K$'s are also unknown,
and thus no definite conclusion can be drawn from the above
calculations as far as the actual values of $f_{_\mrm{NL}}$ are concerned.
Some information on the dominant shapes of non-Gaussianity produced at the bounce
can however be extracted from Eq.~(\ref{eq:fnl}) by making plausible
assumptions on the matrix elements of $\bm P$.
In this paper, we provide two such examples which also highlight the dependence
of the shapes of non-Gaussianities on
the initial conditions at $\eta_-$.

Let us first assume that the functions of the original spectra are
all roughly equal, \ie $K_1 (k_i) K_{1,2}(k_j)\simeq K_3
(k_1,k_2,k_3)$, an approximation that should be roughly valid in many
cosmologically relevant situations. With this simplifying assumption,
one obtains from Eq.~(\ref{eq:fnl}) that 
\begin{equation}
f_{_\mathrm{NL}}\simeq \frac{5}{3\Upsilon} 
\left[ \mathcal{B}(x_2,x_3)-k_1^2\mathcal{C}(x_2,x_3) \right],
\end{equation}
where the dimensionless characteristic shape functions
$ \mathcal{B} $ and $ \mathcal{C} $, which depend only on the ratios
$x_2=k_2/k_1$ and $x_3=k_3/k_1$, are given by
\begin{widetext}
\be
\mathcal{B}(x_2,x_3) \equiv 7+\frac23 \left( \frac{1+x_2^2}{x_3^2} +
\frac{1+x_3^2}{x_2^2} + x_2^2+x_3^2 \right)-3 \left[ \left(
\frac{1-x_2^2}{x_3^2}\right)^2 +
\left( \frac{1-x_3^2}{x_2^2}\right)^2+\left( x_2^2-x_3^2\right)^2\right],
\ee
and
\be
\mathcal{C}(x_2,x_3) = \left(1+x_2+x_3\right) \left(1+x_2-x_3\right)
\left(1+x_3-x_2\right) \times \left(x_2+x_3-1\right)\left(1+
\frac{1}{x_2^2}+\frac{1}{x_3^2}\right).
\ee
These shape functions are displayed in the upper plots of
Fig.~\ref{figure2} where, without loss of generality, we have ordered
the variables by assuming $x_3\leq x_2\leq 1$, with the triangle
inequality given by $x_2-x_3\leq 1\leq x_2+x_3$.  The left-hand plot shows 
the function $\log(|\mcl B|$) and suggests that non-Gaussianities
proportional to $1/\Upsilon$ peak in the folded configuration.
The right-hand plot shows the function $\mcl C$ and suggests 
that non-Gaussianities proportional to the overall factor $k_1^2/\Upsilon$ 
produced in the bouncing phase peak in the
equilateral, take intermediate values in the squeezed, and are small
in the folded configuration.  

Another way to determine the shapes of non-Gaussianities produced in a
bouncing phase in a largely model-independent way consists in assuming
the Bardeen potential to have reached, at $\eta=\eta_-$, the
frozen state characteristic of super-Hubble inflationary evolution,
so that one has $\Psi'\ll\Psi$, leading to
$P_{\Psi'\Psi'}\ll P_{\Psi\Psi'}\ll P_{\Psi\Psi}$. Denoting for
simplicity $P(k_i)\equiv P_{\Psi\Psi}(k_i)$, this then leads to
\begin{align}
f_{_\mathrm{NL}}^\mathrm{frozen}= \frac{180}{243\Upsilon} 
\frac{\mathcal{F}\lsb P(k_1),P(k_2),P(k_3),x_2,x_3\rsb-k_1^2\mathcal{G}\lsb P(k_1),P(k_2),P(k_3),x_2,x_3\rsb}
{P(k_1) P(k_2) +P(k_1) P(k_3) + P(k_2) P(k_3)},
\label{frozen}
\end{align}
where
\begin{align}
&\mathcal{F}\lsb P(k_1),P(k_2),P(k_3),x_2,x_3\rsb = \left[ \frac73 + \frac23\left( \frac{1+x_2^2}{x_3^2}\right)-3 \left( \frac{1-x_2^2}{x_3^2}\right)^2 \right] P(k_1) P(k_2) \nn \\
&\hskip1.4cm + \left[ \frac73 + \frac23\left( \frac{1+x_3^2}{x_2^2}\right)- 3 \left( \frac{1-x_3^2}{x_2^2}\right)^2 \right] P(k_1) P(k_3) + \left[ \frac73 + \frac23\left( x_2^2+x_3^2\right)-3 \left( x_2^2 - x_3^2\right)^2 \right] P(k_2) P(k_3),&
\end{align}
and 
\begin{align}
\mathcal{G}\lsb P(k_1),P(k_2),P(k_3),x_2,x_3\rsb  = &\left(1+x_2+x_3\right) 
\left(1+x_2-x_3\right) 
\left(1+x_3-x_2\right)
\left(x_2+x_3-1\right)\nn\\
& \times \left[\frac{P(k_1)P(k_2)}{x_3^2} + \frac{P(k_1)P(k_3)}{x_2^2} + P(k_2)P(k_3)\right]\,.
\end{align}

In order to go one step further and actually
evaluate the non-Gaussianities produced during
the contraction-to-expansion transition, we assume,
as is often done, that the spectrum produced during
the contraction phase not only passed through the bounce
unchanged but also that it is in agreement with the data.
Assuming observational constraints to be those of
P{\footnotesize LANCK}, we obtain that, in our notations,
this requires the power spectrum to behave as a power
law $P(k) \propto k^{n_\mathrm{s}-4}$, with \cite{Ade:2013uln}
$n_\mathrm{s} = 0.9603 \pm 0.0073$. The ratios of
power spectra in Eq.~(\ref{frozen}) then read
\begin{align}
\frac{P(k_1)P(k_2)}{P(k_1) P(k_2) +P(k_1) P(k_3) + P(k_2) P(k_3)} &= \left[ 
1+ \left(\frac{x_3}{x_2}\right)^{n_\mathrm{s}-4}+x_3^{n_\mathrm{s}-4}
\right]^{-1},\\
\frac{P(k_1)P(k_3)}{P(k_1) P(k_2) +P(k_1) P(k_3) + P(k_2) P(k_3)} &= \left[ 
1+ \left(\frac{x_2}{x_3}\right)^{n_\mathrm{s}-4}+x_2^{n_\mathrm{s}-4}
\right]^{-1},\\
\frac{P(k_2)P(k_3)}{P(k_1) P(k_2) +P(k_1) P(k_3) + P(k_2) P(k_3)} &= \left[ 
1+ \left(x_2\right)^{4-n_\mathrm{s}}+\left(x_3\right)^{4-n_\mathrm{s}}
\right]^{-1},
\end{align}
\end{widetext}
The shape functions that can be formed by combining Eqs~(20) to (24) 
are shown in the lower panel of Fig.~\ref{figure2}. In this case, 
the equilateral configuration is favored while both the squeezed and 
folded configurations are subdominant.

To conclude, let us discuss two interesting limiting behaviors of the
model. The first is the quasi-de Sitter approximation which, as
mentioned before, is equivalent to having $\Upsilon\ll 1$. In this
limit, and contrary to the single field slow-roll inflationary
situation, Eqs.~(\ref{fNLequiv}-\ref{fNLfold}) show that large amounts
of non-Gaussianities are produced in all possible shapes, with
$f_{_\mrm{NL}}\propto \Upsilon^{-1}\gg 1$.  Thus, although large
non-Gaussianities in inflation often stem from a violation of slow
roll, in the bouncing case, the closer one is to a de Sitter bounce,
the more non-Gaussianities are produced.  The second limiting behavior
is perhaps more relevant for comparison with observational data, as it
is not based on any prerequisite regarding the structure of the
bounce.  As seen from Eqs.~(\ref{fNLequiv}) to (\ref{fNLfold}), the
parameter $f_{_\mrm{NL}}$ is scale dependent, and in particular, is
proportional to $k^2$ in the equilateral and squeezed
configurations. In a cosmological background with closed spatial
sections and with a present value of $\Omega_{\mcl K}$ of the order of
$10^{-2}$, the mode numbers are, as discussed above, in the range
$\left[ 10^2,10^8\right]$, so the expected non-Gaussianities are
predicted to be extremely large right after the bouncing phase.  In
both limits, the amount of non-Gaussianity produced greatly exceeds
the current observational limits and the validity of the perturbative
expansion may be brought into question. We conjecture that this is
likely to be a generic and potentially serious problem for
nonsingular bouncing cosmologies.

\acknowledgements{{\it Acknowledgements.}~M.~L.~was supported by a
  South African NRF reseach grant while this research was conducted. X.~G.~is supported by a JSPS Grant-in-Aid for Scientific Research No.~25287054. The authors thank J\'er\^ome Martin for useful comments.}

\bibliography{references}

\end{document}